\documentclass[12pt]{article}
\usepackage{graphicx}
\usepackage{amsmath}

\newtheorem{theorem}{Theorem}
\newtheorem{acknowledgement}[theorem]{Acknowledgement}

\input{tcilatex}

\begin{document}

\title{Correlation Functions of The Tri-critical 3-States Potts Model}
\author{\textsc{S. Balaska and K. Demmouche} \\
{\small Laboratoire de Physique Th\'{e}orique. Universitiy of Oran
Es-S\'{e}nia. \ }\\
{\small \ 31100 Oran Es-S\'{e}nia. Algeria}}
\maketitle

\begin{abstract}
We build the Z$_{3}$ invariants fusion rules associated to the (D$_{4}$,A$%
_{6}$) conformal algebra. This algebra is known \ to describe the
tri-critical Potts model. The 4-pt correlation functions of critical fields
are developed in the bootstrap approach, and in the other hand, they are
written in term of integral representation of the conformal blocks. By
comparing both the expressions, one can determine the structure constantes
of the operator algebra.
\end{abstract}

\section{Introduction}

The Conformal Field Theories on 2d euclidean spaces are of major importance,
they allow through their mathematical tools the investigation of several
topics in physics, such as statistical physics in condensed matter and
String theory in high-energy physics. The minimal models of central charge $%
c=1-6/m(m+1)$ ($m=6$ in our case ) form a particular class of conformal
theories, they describe the critical (or tricritical) points of known 2d
statistical systems, they were introduced in the seminal work of Belavin and 
\textit{al} \cite{bpz}. Just after that discovery, Dotsenko calculated the
linear differential equations obeyed by correlators containing some
degenerate conformal fields, and he applied those results to the $Z_{3}$
Potts model \cite{Z3potts}. This was done before the discovery of the
modular ADE classification \cite{CIZ}. Next, the connection was been made in
many ways such in \cite{pasquier}.

Recently, the description of systems with tricritical point by non-minimal
conformal theories was established in \cite{Dots-recent}.

Our interest here is the determination of the structure constants of the
tri-critical Potts algebra from that of the (D$_{4}$,A$_{6}$) one. For that
we construct the $Z_{3}$ invariant fusion rules and use a method which
consists in comparing the integral representation of the correlation
functions with their developments obtained in the bootstrap approach. This
method was first used by McCabe in \cite{Mcb-w-s}, \cite{MCABE} to determine
the structure constants of the critical 3-states Potts model .

Because the studied algebra here contains an integer spin field $W_{5}$
which generates an extended algebra, the model become classified in a
diagonal series, and it is possible to study such models in the context of
extended theory \cite{Fat-Zam}, \cite{w-symetry}.

As done for the associated conformal algebra of the critical 3-states Potts
model, the identification in this work for the \textit{tri}-critical
3-states Potts model is done by writing the operators of spin density (order
parameters ) as a certain complex function of the primary operators of the (D%
$_{4}$,A$_{6}$) algebra studied in \cite{BK}. Knowing that the Hamiltonian
of the critical model admits a Z$_{3}$ symmetry, one can build the fusion
rules from those established in \cite{BK} where the action of the
transformation T$_{2}$ was considered. It results from this fact that the
obtained fusion rules are naturally Z$_{3}$ invariants.

The paper is organized as follows, in section 1 we identify the operator
algebra of the tri-critical 3-states Potts medel from that of the (D$_{4}$,A$%
_{6}$) model. In section 2, we establish the Z$_{3}$ invariant fusion rules.
The section 3 and 4 are devoted to write the conformal blocks of the
correlation functions in their integral representation and the equations
obtained from the bootstrap approach. Finally, we conclude by giving the
numerical values of the structure constants.

\section{The tricritical Potts model from the (D$_{4}$,A$_{6}$) model}

The (D$_{4}$,A$_{6}$) model is the complementary series corresponding to the
minimal model $\mathcal{M}(7,6)$, it contains basic operators having
conformal weights $h_{r,s}$ which are given by Kac formula as follows

\bigskip

\begin{center}
s$\uparrow $ \ \ 
\begin{tabular}{|c|c|c|c|c|}
\hline
&  &  &  &  \\ \hline
&  &  &  & 1/7 \\ \hline
&  &  & 5/56 & 5/7 \\ \hline
&  & \textbf{1/21} & 33/56 & 12/7 \\ \hline
& 1/56 & 10/21 & 85/56 & 22/7 \\ \hline
0 & 3/8 & 4/3 & 23/8 & 5 \\ \hline
\end{tabular}

\ \ \ \ \ \ \ \ \ \ \ \ \ \ \ \ \ \ \ \ \ \ \ \ \ \ \ \ \ \ \ \ \ r$%
\rightarrow $

\textbf{Table 1}: Conformal grid (Kac table: $h_{r,s}$ )

\bigskip
\end{center}

The physical Hilbert space of the (D$_{4},$A$_{6}$) conformal model is given
by the modular invariant partition function hereafter \cite{CIZ} 
\begin{eqnarray}
\mathcal{Z}_{(D_{4},A_{6})} &=&\chi _{1,1}\overline{\chi }_{1,1}+\chi _{1,3}%
\overline{\chi }_{1,3}+\chi _{1,5}\overline{\chi }_{1,5}+\chi _{1,1}%
\overline{\chi }_{5,1}+\chi _{1,3}\overline{\chi }_{5,3}+\chi _{1,5}%
\overline{\chi }_{5,5}  \notag \\
&&+\chi _{5,1}\overline{\chi }_{1,1}+\chi _{5,3}\overline{\chi }_{1,3}+\chi
_{5,5}\overline{\chi }_{1,5}+\chi _{5,1}\overline{\chi }_{5,1}+\chi _{5,3}%
\overline{\chi }_{5,3}+\chi _{5,5}\overline{\chi }_{5,5}  \notag \\
&&+2\chi _{3,1}\overline{\chi }_{3,1}+2\chi _{3,3}\overline{\chi }%
_{3,3}+2\chi _{3,5}\overline{\chi }_{3,5}  \label{D4}
\end{eqnarray}
where $\chi _{rs},$ ($\overline{\chi }_{\overline{r}\overline{s}})$ are
characters of representations of the left (right) Virasoro algebra of
central charge $c=\frac{7}{6}$. We note that not all the fields of the $%
\mathcal{M}(7,6)\ $minimal model are present , the existence of spinless as
well as spin left-right combinations and finally the presence of two copies
for each of the spinless combinations $\Phi _{(3,s|3,s)}$\medskip $=\phi
_{(3,s)}\otimes \overline{\phi }_{(3,s)}$ \ for $s=1,2,3$. The conformal
anomaly $c$ in the context of critical phenomena can be interpreted as the
ground state energy of the statistical model, it is also related to the
magnitude of the Casimir effect in the field theory \cite
{affleck-nightingale}.

Long time ago, Friedan et \textit{al} \cite{FQS}\ gave some conformal fields
included in the tricritical Potts model. Some of those appear in (\ref{D4}).
There are several manners to identify primary field living in the conformal
model which correspond to the spin density operator (order parameter) of the
critical model.

The identification is made by the knowledge of critical exponents $x_{\Phi }$
(magnetic $x_{H}$ and thermal $x_{T}$) given in statistical physics \cite
{ABF}. Those exponents are related to the conformal dimensions $h$ of
primaries by the equations \cite{HENKEL} 
\begin{equation}
\left\{ 
\begin{array}{c}
x_{\epsilon }=x_{T}=d-y_{T} \\ 
x_{\sigma }=x_{H}=d-y_{H}
\end{array}
\right\}
\end{equation}
with $x_{\Phi }=h_{\Phi }+\overline{h}_{\Phi }$, and the exponents $y_{T}$
and $y_{H}$ are given in many references such the conjecture of Baxter \cite
{nienhuis}, \cite{nigs}.

From where we obtain for the spin $\sigma $\ and energy $\epsilon $
operators 
\begin{eqnarray}
h_{\epsilon } &=&\frac{1}{7}  \notag \\
h_{\sigma } &=&\frac{1}{21}
\end{eqnarray}

Thus, the remaining operators are identified in the same way. Finally, the
complete set of critical fields is shown in table 2,

\begin{center}
\begin{tabular}{|l|l|l|}
\hline
Primary fields & $\Delta =h+\overline{h}$ & Critical fields \\ \hline
$\Phi _{(33\mid 33)}$ & \textbf{2/21} & $\sigma $ \\ 
$\Phi _{(32\mid 32)}$ & 20/21 & $\sigma ^{\prime }$ \\ 
$\Phi _{(31\mid 31)}$ & 8/3 & $\sigma "$ \\ 
$\Phi _{(13\mid 13)}$ & 10/7 & $\epsilon $ \\ 
$\Phi _{(12\mid 12)}$ & 2/7 & $\epsilon ^{\prime }$ \\ 
$\Phi _{(51\mid 11)}$ & 5 & $\epsilon "$ \\ 
$\Phi _{(15\mid 12)}$ & 23/7 & $\Phi _{(15\mid 12)}$ \\ 
$\Phi _{(14\mid 13)}$ & 17/7 & $\Phi _{(14\mid 13)}$ \\ \hline
\end{tabular}

\bigskip

\textbf{Table 2}: Critical fields and the corresponding primaries

\bigskip
\end{center}

At this stage, we would like to remark that the spin operator is again in
the center of the conformal grid (Table 1), this fact was already predicted
by Fridan \cite{FQS} and used by Dotsenko \cite{Z3potts} to obtain the
conformal algebra of the Potts model.

Another way to identify the conformal field $\phi _{3.3}$ as the spin
operator consists in using the relations between the exactly known critical
exponents $\delta $ and $\eta $ (magnetic exponent) of the tricritical Potts
model and the conformal weight of the primary field corresponding to the
spin operator. The critical exponent $\delta $ is related to $\eta ,$ from
the scaling laws, by

\begin{equation}
\delta =\frac{d+2-\eta }{d-2+\eta }
\end{equation}

$(d=2)$ in our case.

At the same time $\eta $ is the exponent appearing in spin 2-pt correlations
in the statistical model and it is related to the conformal weight by

\begin{equation}
\eta =4h_{r,s}
\end{equation}
using the results of Baxter \cite{nienhuis} we have $\delta =20$ and we
obtain

\begin{equation}
\eta =\frac{4}{21}
\end{equation}

Finally, the conformal weight of the spin operator is, as already shown, $%
h_{3.3}=h_{3.4}=\frac{1}{21}.$

\section{The Z$_{3}$ invariants fusion rules}

In the physical model, the Hamiltonian of the Potts model is invariant under
the Z$_{3}$ symmetry. The spin density operator is sensitive to this action,
whereas $\epsilon $\ remains invariant. Therefore, if one would like to
define the action of Z$_{3}$ on the operator with non vanishing spin ($s$),
it is sufficient to observe that the \textit{left-right} components of such
operator come from the thermal algebra ($\epsilon $), and consequently we
conclude by saying that those operators are Z$_{3}$ invariants.

Now, we are interesting in the construction of the fusion rules between the
critical fields of the model. Before that, it is necessary to note that ,
for example, the field $\Phi _{(33\mid 33)}$ representing the critical field 
$\sigma $ is doubly degenerate \cite{BK}. This fact required the
introduction of the Z$_{2}$ symmetry which separates between the two copies
noted $\Phi _{(33\mid 33)}^{+}$ and $\Phi _{(33\mid 33)}^{-}$. For
representing the operator $\sigma ,$we realize the following change of basis 
\begin{eqnarray}
\sigma &=&\frac{1}{\sqrt{2}}\left( \Phi _{(33\mid 33)}^{+}+i\Phi _{(33\mid
33)}^{-}\right)  \notag \\
\overline{\sigma } &=&\frac{1}{\sqrt{2}}\left( \Phi _{(33\mid 33)}^{+}-i\Phi
_{(33\mid 33)}^{-}\right)  \label{eq3}
\end{eqnarray}
in this case, the action Z$_{3}$ is defined as follows 
\begin{eqnarray}
Z_{3} &:&\sigma \rightarrow e^{i\frac{2\pi }{3}}\sigma  \notag \\
Z_{3} &:&\overline{\sigma }\rightarrow e^{-i\frac{2\pi }{3}}\overline{\sigma 
}  \label{change-basis}
\end{eqnarray}

Now to derive the Z$_{3}$ invariant fusion rules between the different
fields of the model, we need to use the Z$_{2}$ invariant ones obtained in 
\cite{BK} with respect to the transformation of type T$_{2}.$ This
transformation acts as follows: 
\begin{eqnarray}
&&\Phi \text{ \ \ }\underrightarrow{\text{T}_{2}}\text{ \ \ \ }\left\{ 
\begin{array}{c}
\Phi \text{ \ if \ }s_{\Phi }=0 \\ 
-\Phi \text{ \ if \ }s_{\Phi }\neq 0
\end{array}
\right\}  \notag \\
&&\Phi ^{\pm }\text{ \ \ }\underrightarrow{\text{T}_{2}}\text{ \ \ \ }\pm
\Phi ^{\pm }\text{ }
\end{eqnarray}
where $s_{\Phi }=h-\overline{h}$, is the conformal spin of the field.

This leads to the following new fusion rules 
\begin{eqnarray}
\sigma .\sigma &=&\overline{\sigma }+\overline{\sigma }^{\prime }+\overline{%
\sigma }"  \notag \\
\sigma .\overline{\sigma } &=&1+\epsilon +\epsilon ^{\prime }+(51\mid
51)+(14\mid 14)+(15\mid 15)+(51\mid 11)+(11\mid 51)  \notag \\
+(13 &\mid &14)+(14\mid 13)+(15\mid 12)+(12\mid 15)  \notag \\
\overline{\sigma }.\overline{\sigma } &=&\sigma +\sigma ^{\prime }+\sigma " 
\notag \\
\epsilon .\epsilon &=&1+\epsilon ^{\prime }  \label{eq4} \\
\epsilon .\sigma &=&\sigma +\sigma ^{\prime }  \notag \\
\epsilon .\overline{\sigma } &=&\overline{\sigma }+\overline{\sigma }%
^{\prime }  \notag \\
\sigma ".\sigma " &=&\overline{\sigma }"  \notag \\
\sigma ".\overline{\sigma }" &=&1+(51\mid 51)+(11\mid 51)+(51\mid 11)  \notag
\end{eqnarray}

It should be noted here that the transformations T$_{1}$, T$_{3}$ and T$_{4}$
figuring in \cite{BK} with the basis change (\ref{eq3}) do not lead to a Z$%
_{3}$ invariant fusion rules contrary to the case (\ref{eq4}). From the
obtained fusion rules, one can first note that the product spin density-spin
density generates all the conformal algebra of the model and we see that
only the operators of the first, the thirth and the fifth column of table
(1) contribute ( couple to $\sigma $) to the operator product of spins.

\section{Correlation functions and conformal blocks}

Let us first recall a fundamental property of the structure constants, which
rises from the radial quantification of conformal theories. The Operator
Product Expansion (OPE) of two conformal fields is defined by \cite{bpz}, 
\begin{equation}
\Phi _{i}(z,\overline{z}).\Phi _{j}(0,0)=\sum_{k}\frac{\widetilde{C}%
_{ij}^{k}\Phi _{k}(0,0)}{z^{h_{i}+h_{j}-h_{k}}\times cc}  \label{eq0}
\end{equation}

Where $h,\overline{h}$ are the conformal weights. In the other hand, the
3-pt correlation function is fixed by conformal invariance and has the
following form 
\begin{equation}
\left\langle \Phi _{i}(1)\Phi _{j}(2)\Phi _{k}(3)\right\rangle =\frac{C_{ijk}%
}{%
z_{12}^{h_{i}+h_{j}-h_{k}}z_{13}^{h_{i}+h_{k}-h_{j}}z_{23}^{h_{j}+h_{k}-h_{i}}\times cc%
}  \label{eq1}
\end{equation}

The $\widetilde{C}_{ij}^{k}$\ are the structure constants and they are
related to the coefficients $C_{ijk}$ appearing in the (\ref{eq1}) by 
\begin{equation}
C_{ijk}=(-)^{S_{k}}\widetilde{C}_{ij}^{k}  \label{eq2}
\end{equation}

The relation (\ref{eq2}) becomes, 
\begin{equation}
C_{abc}=(-)^{S_{c}}\widetilde{C}_{ab}^{\overline{c}}
\end{equation}
when $a,b$ and $c$ indicate the critical fields constructed from the complex
combination (\ref{change-basis}). $S_{k}=h-\overline{h}$ is the spin of the
field $\Phi _{k}$ . Our purpose here is the determination of these universal
quantities in the case of the tricritical Potts algebra.

The 4-pt conformal correlations are composed of two chiral parts (holomophic
and anti-holomorphic), coupled with an unspecified non diagonal\footnote{%
This is because the existence in the $D-series$ of non diagonal primary
fields with non vanishing spin, such as $\Phi _{(51\mid 11)}$.} coupling
constants. In our case, with the choosen correlation function, we write 
\begin{equation}
\left\langle \Phi _{(kl,\overline{k}\overline{l})}\sigma \overline{\sigma }%
\Phi _{(kl,\overline{k}\overline{l})}\right\rangle =\sum_{n,m}\gamma _{nm}%
\mathcal{F}_{n}\overline{\mathcal{F}}_{m}  \label{cor-fun}
\end{equation}
where $\mathcal{F}_{n}$\ are the conformal blocks, and \ $\gamma _{nm}$ are
the coupling constants which are not diagonal in the case of (D$_{4}$,A$_{6}$%
) conformal model, we will concentrate on the holomorphic part (all is true
for the antiholomophic one).

The conformal blocks are given by a correlation of vertex operators
(integral representation) \cite{PETKV}, 
\begin{equation}
\mathcal{F}_{k_{1}k_{2}}^{(33)}=\left\langle \mathcal{V}_{\alpha _{kl}}(1)%
\mathcal{V}_{\alpha _{33}}(2)\mathcal{V}_{\alpha _{33}}(3)\mathcal{V}%
_{2\alpha _{0}-\alpha _{kl}}(4)\mathcal{Q}_{-}^{N}\mathcal{Q}%
_{+}^{M}\right\rangle  \label{eq5}
\end{equation}

$\mathcal{V}_{\alpha }$ are the vertex operator of charges $\alpha _{rs}=%
\frac{1}{2}[(1-r)\alpha _{-}+(1-s)\alpha _{+}]$, where $r,s$ belongs to the
Kac indices. The screening operators $\mathcal{Q}_{\pm }$ are injected into
the correlation function so that the neutrality condition $\sum_{i}\alpha
_{i}=2\alpha _{0}$ is verified, we have 
\begin{equation}
\left\langle \prod_{i=1}^{K}V_{\alpha _{i}}(z_{i})\right\rangle =\left\{ 
\begin{array}{c}
\prod_{i<j}^{K}(z_{i}-z_{j})^{2\alpha _{i}\alpha _{j}}\;\;\;\;\;\text{if \ \
\ \ \ \ \ \ }\sum_{i=1}^{K}\alpha _{i}=2\alpha _{0} \\ 
\\ 
0\;\ \ \ \ \ \ \ \ \ \ \ \ \ \ \ \ \ \ \ \ \ \ \ \text{otherwise}
\end{array}
\right\}
\end{equation}

In the coulomb gas formalism \cite{DF}, the operators $\mathcal{Q}_{\pm }$
are integrals over closed contour of primary field of conformal weight $h=1$.

The neutrality condition applied to the conformal block in (\ref{eq5}) gives 
$N=M=2$, then it becomes 
\begin{eqnarray}
\mathcal{F}_{k_{1}k_{2}}^{(33)} &=&f(\eta )\lambda _{k_{1}}(\rho ^{\prime
})\lambda _{k_{2}}(\rho )\Im _{k_{1}k_{2}}^{(33)}(a,b,c;\rho ;z)
\label{bloc} \\
k_{1},k_{2} &=&1,2,3  \notag
\end{eqnarray}
with the cross ration $\eta =\frac{z_{12}z_{34}}{z_{13}z_{24}}$, and $%
\lambda $ is an appropriate factor, $f$ is function of $\eta $. The $\Im $'s
are integrals over closed contours with ($k_{1}\times k_{2}$) independent
solutions having monodromy properties, this means that they admit a
development around $\eta \rightarrow 0$ et $1-\eta \rightarrow 0$. The
integrals $\Im $'s depend of certain exponents $a,b,c,\rho $ , where their
associated values in the case of ($kl,\overline{k}\overline{l}=33,33$) being 
\begin{eqnarray}
a &=&b=c=\frac{-1}{3}  \notag \\
a^{\prime } &=&b^{\prime }=c^{\prime }=\frac{2}{7} \\
\rho ^{\prime } &=&\frac{1}{\rho }=\frac{6}{7}  \notag
\end{eqnarray}

Now, in the limit $\eta \rightarrow 0$ ($s-$channel), the conformal block
has the following form 
\begin{equation}
\mathcal{\Im }_{k_{1}k_{2}}^{(33)}=N_{k_{1}k_{2}}^{(33)}\eta ^{K}[1+\mathcal{%
O}(\eta )]
\end{equation}
with 
\begin{eqnarray*}
K &=&(k_{1}-1)[1+a^{\prime }+c^{\prime }+\rho ^{\prime }(k_{1}-2)] \\
&&+(k_{2}-1)[1+a+c+\rho (k_{2}-2)]-2(k_{1}-1)(k_{2}-1)
\end{eqnarray*}

The $N$'s are the normalization constants of the integrals $\mathcal{\Im }$%
's \cite{DF}. So, there are nine normalized conformal blocks which can be
written in a matrix form as 
\begin{equation}
\mathcal{F}_{k_{1}k_{2}}^{(33)}\approx \left( 
\begin{array}{lll}
\eta ^{1/21} & \eta ^{8/21} & \eta ^{64/21} \\ 
\eta ^{34/21} & \eta ^{-1/21} & \eta ^{13/21} \\ 
\eta ^{103/21} & \eta ^{26/21} & \eta ^{-2/21}
\end{array}
\right)  \label{cb-s}
\end{equation}

In the other hand, in the $t-$channel ($1-\eta \rightarrow 0$), the
conformal blocks are written via a monodromy transformation as follow 
\begin{equation}
\Im _{k_{1}k_{2}}^{(33)}(a,b,c;\rho ;\eta )=\sum_{kk^{\prime }}\Psi
_{(k_{1}k_{2},kk^{\prime })}\widetilde{\Im }_{kk^{\prime
}}^{(33)}(b,a,c;\rho ;1-\eta )  \label{cb-t}
\end{equation}
where the $\Psi _{(k_{1}k_{2},kk^{\prime })}=\psi _{k_{1}k}^{\prime }\times
\psi _{k_{2}k^{\prime }}$ are the elements of the monodromy matrix.

Finally, the holomorphic and the anti-holomorphic conformal blocks are
combined by the coupling $\gamma $ to build the correlation function (\ref
{cor-fun}).

\section{Bootstrap equations}

The operator algebra (OPA) in (\ref{eq0}) is associative, which induced a
crossing symmetry of the 4-pt correlation functions, this fact allows to
write the bootstrap equations \cite{bpz}. For the correlation function 
\begin{equation}
G=\left\langle \Phi _{1}(z_{1},\overline{z}_{1})\Phi _{2}(z_{2},\overline{z}%
_{2})\Phi _{3}(z_{3},\overline{z}_{3})\Phi _{4}(z_{4},\overline{z}%
_{4})\right\rangle
\end{equation}
,we have in the $s-$channel ($z_{12},z_{34}\rightarrow 0$)

\begin{equation}
G(z_{1},...,\overline{z}_{4})=\sum_{m}\frac{(-1)^{s_{m}}\widetilde{C}%
_{12}^{m}\widetilde{C}_{34}^{m}\left[ 1+\mathcal{O}(z_{12},\overline{z}_{34})%
\right] }{z_{12}^{h_{1}+h_{2}-h_{m}}z_{34}^{h_{3}+h_{4}-h_{m}}z_{24}^{2h_{m}}%
\overline{z}_{12}^{\overline{h}_{1}+\overline{h}_{2}-\overline{h}_{m}}%
\overline{z}_{34}^{\overline{h}_{3}+\overline{h}_{4}-\overline{h}_{m}}%
\overline{z}_{24}^{2\overline{h}_{m}}}  \label{S-}
\end{equation}
and in the $t-$channel ($z_{14},z_{23}\rightarrow 0$)

\begin{equation}
G(z_{1},...,\overline{z}_{4})=\sum_{n}\frac{(-1)^{s_{n}}\widetilde{C}%
_{14}^{n}\widetilde{C}_{23}^{n}\left[ 1+\mathcal{O}(z_{14},\overline{z}_{23})%
\right] }{z_{41}^{h_{1}+h_{4}-h_{n}}z_{23}^{h_{3}+h_{2}-h_{n}}z_{13}^{2h_{m}}%
\overline{z}_{41}^{\overline{h}_{1}+\overline{h}_{4}-\overline{h}_{n}}%
\overline{z}_{23}^{\overline{h}_{3}+\overline{h}_{2}-\overline{h}_{n}}%
\overline{z}_{13}^{2\overline{h}_{n}}}  \label{T-}
\end{equation}

Where $m$ and $n$ belong to the algebra (\ref{eq4}). The appearance of the
structure constants is realized in these two last expressions. To determine
them, one has to choose the appropriate correlation function where the
desired structure constant appear. Then one begin by writing the
corresponding bootstrap equations and next write the correlation function in
term of conformal blocks in their limits to compare them with the bootstraps.

For the correlation function 
\begin{equation}
G_{1}=\left\langle \sigma \sigma \overline{\sigma }\overline{\sigma }%
\right\rangle
\end{equation}
where we calculated the corresponding conformal blocks in (\ref{bloc}), it
is easy to write the bootstrap equations using de fusion rules (\ref{eq4}),
and we have in the $s-$ and the $t-$channel respectively 
\begin{mathletters}
\begin{eqnarray}
G_{1} &\sim &\frac{\left| \widetilde{C}_{\sigma \sigma }^{\overline{\sigma }%
}\right| ^{2}}{\left| z_{12}z_{34}\right| ^{2/21}\left| z_{24}\right| ^{4/21}%
}+\frac{\left| \widetilde{C}_{\sigma \sigma }^{\overline{\sigma }^{\prime
}}\right| ^{2}}{\left| z_{12}z_{34}\right| ^{-16/21}\left| z_{24}\right|
^{40/21}}+\frac{\left| \widetilde{C}_{\sigma \sigma }^{\overline{\sigma }%
"}\right| ^{2}}{\left| z_{12}z_{34}\right| ^{52/21}\left| z_{24}\right|
^{16/3}}+..  \label{eq-s} \\
G_{1} &\sim &\frac{1}{\left| z_{14}z_{23}\right| ^{4/21}}+\frac{\left| 
\widetilde{C}_{\sigma \overline{\sigma }}^{\epsilon }\right| ^{2}}{\left|
z_{14}z_{23}\right| ^{-2/21}\left| z_{13}\right| ^{4/7}}+\frac{\left| 
\widetilde{C}_{\sigma \overline{\sigma }}^{\epsilon ^{\prime }}\right| ^{2}}{%
\left| z_{14}z_{23}\right| ^{26/21}\left| z_{13}\right| ^{20/7}}+..
\label{eq-t}
\end{eqnarray}

Now, when comparing (\ref{eq-s}) with the limit form of the conformal blocks
(\ref{cb-s}), one obtains the following identities 
\end{mathletters}
\begin{eqnarray}
\gamma _{(22,22)}\left( N_{22}^{(33)}\right) ^{2} &\approx &\left| 
\widetilde{C}_{\sigma \sigma }^{\overline{\sigma }}\right| ^{2}  \notag \\
\gamma _{(12,12)}\left( N_{12}^{(33)}\right) ^{2} &\approx &\left| 
\widetilde{C}_{\sigma \sigma }^{\overline{\sigma }^{\prime }}\right| ^{2}
\label{sys-1} \\
\gamma _{(32,32)}\left( N_{32}^{(33)}\right) ^{2} &\approx &\left| 
\widetilde{C}_{\sigma \sigma }^{\overline{\sigma }"}\right| ^{2}  \notag
\end{eqnarray}
and the remaining coupling constants $\gamma $\ are nulls.

The second expression (\ref{eq-t}) will be compared with the development of
the conformal blocks (\ref{cb-t}) when $1-\eta \rightarrow 0$ . The
correlation function written in this case has the form 
\begin{equation}
G_{1}=\limfunc{Tr}\overline{\widetilde{\mathcal{F}}}^{T}\left( \Psi
^{T}\gamma \Psi \right) \widetilde{\mathcal{F}}  \label{psi-psi}
\end{equation}
thus, when injecting (\ref{cb-t}) in the last expression, it leads to the
following equations for the first three terms 
\begin{eqnarray}
\left( \gamma _{(12,12)}\Psi _{(12,33)}^{2}+\gamma _{(22,22)}\Psi
_{(22,33)}^{2}+\gamma _{(32,32)}\Psi _{(32,33)}^{2}\right) \lambda
_{3}^{2}\lambda _{3}^{2}\widetilde{N}_{33}^{2} &=&1  \notag \\
\left( \gamma _{(12,12)}\Psi _{(12,11)}^{2}+\gamma _{(22,22)}\Psi
_{(22,11)}^{2}+\gamma _{(32,32)}\Psi _{(32,11)}^{2}\right) \lambda
_{1}^{2}\lambda _{1}^{2}\widetilde{N}_{11}^{2} &=&\left| \widetilde{C}%
_{\sigma \overline{\sigma }}^{\epsilon }\right| ^{2}  \notag \\
\left( \gamma _{(12,12)}\Psi _{(12,23)}^{2}+\gamma _{(22,22)}\Psi
_{(22,23)}^{2}+\gamma _{(32,32)}\Psi _{(32,23)}^{2}\right) \lambda
_{2}^{2}\lambda _{3}^{2}\widetilde{N}_{23}^{2} &=&\left| \widetilde{C}%
_{\sigma \overline{\sigma }}^{\epsilon ^{\prime }}\right| ^{2}  \notag \\
&&  \label{sys-2}
\end{eqnarray}

It is noted here that before solving the system of equations (\ref{sys-1})
and (\ref{sys-2}), one can reduce the number of unknown coupling constants
by calculating them from other correlation. For example, the constants in (%
\ref{sys-2}) can be obtained from the correlation $\left\langle \sigma
\epsilon \epsilon \overline{\sigma }\right\rangle $.

In (\ref{psi-psi}) there are three vanishing coupling $\widetilde{\gamma }%
_{ij,ij}$, this is because the presence of non-diagonal fields in the fusion 
$\sigma .\overline{\sigma },$ and more precisely it concerns the coupling $%
\widetilde{\gamma }_{12,12}$, $\widetilde{\gamma }_{22,22}$ and $\widetilde{%
\gamma }_{32,32}$. Hence, we have the following algebraic equation 
\begin{equation}
\sum_{(\mu \nu )}\gamma _{\mu \nu ,\mu \nu }\psi _{(\mu \nu )k}^{\prime
}\psi _{(\mu \nu )l}^{\prime }=0\text{ \ if }k\neq l  \label{mine}
\end{equation}

This equation allows to write the ratio between the constants $\gamma $%
.These obtained ration are injected in the equations (\ref{sys-1}) and (\ref
{sys-2}) which lead at the end to the determination of \ the values of the
structure constants.

\section{Conclusion}

In this section, and after calculation the numerical values of the
normalization constants $N_{ij}$, the monodromy elements $\Psi _{\alpha
\beta ,\mu \nu }$ and the pre-factors $\lambda _{k}$ ,we resumed the
obtained numerical values of the structure constants in table 3.

\begin{center}
\bigskip \bigskip 
\begin{tabular}{|l|l|}
\hline
Structure constante & Value \\ \hline
$\widetilde{C}_{\sigma \sigma }^{\overline{\sigma }}$ & $\left( \frac{B}{%
\theta }\right) ^{1/2}\frac{1}{\sqrt{3}}\frac{s(1)}{s(2)}\frac{275}{588}%
\frac{\Gamma ^{2}(\frac{5}{6})\Gamma (\frac{1}{7})\Gamma (-\frac{1}{7})}{%
\sqrt{\pi }\Gamma (\frac{2}{7})\Gamma (\frac{4}{7})\Gamma (\frac{6}{7})}$ \\ 
\hline
$\widetilde{C}_{\sigma \sigma }^{\overline{\sigma }^{\prime }}$ & $\left( 
\frac{A}{\theta }\right) ^{1/2}\frac{1}{\sqrt{3}}\frac{55}{4032}\frac{\Gamma
^{2}(\frac{5}{6})\Gamma (\frac{1}{7})\Gamma (-\frac{1}{7})\Gamma (\frac{5}{7}%
)}{\sqrt{\pi }\Gamma (\frac{3}{7})\Gamma (\frac{6}{7})}$ \\ \hline
$\widetilde{C}_{\sigma \sigma }^{\overline{\sigma }"}$ & $\left( \frac{1}{%
\theta }\right) ^{1/2}\frac{s(2)}{s(1)}\frac{55}{81}\frac{\Gamma ^{2}(\frac{5%
}{6})}{\sqrt{\pi }}$ \\ \hline
$\widetilde{C}_{\sigma \overline{\sigma }}^{\epsilon }$ & $\left( \frac{%
\theta _{1}}{\theta }\right) ^{1/2}\frac{1520640}{16807}\frac{\Gamma ^{2}(%
\frac{5}{6})\Gamma ^{2}(\frac{1}{7})\Gamma (-\frac{5}{7})\Gamma (-\frac{8}{7}%
)\Gamma (-\frac{2}{7})}{\Gamma ^{2}(\frac{2}{3})\Gamma ^{2}(\frac{6}{7}%
)\Gamma (\frac{3}{7})\Gamma (\frac{2}{7})}$ \\ \hline
$\widetilde{C}_{\sigma \overline{\sigma }}^{\epsilon ^{\prime }}$ & $\left( 
\frac{\theta _{2}}{\theta }\right) ^{1/2}\frac{s(1)}{s(2)}\frac{4}{77}\frac{%
\Gamma (\frac{1}{7})}{\Gamma (\frac{2}{7})\Gamma (\frac{4}{7})}$ \\ \hline
\end{tabular}

\textbf{Table 3}: Strucrure constants of the Tri-critical Potts model
\end{center}

With, 
\begin{eqnarray*}
A &=&2\frac{s^{2}(3)}{s^{2}(1)}\frac{1}{s(1)s(2)-s(3)}\text{ ; }B=\frac{s(3)%
}{s^{2}(1)}\text{ ; }s(x)=\sin (\frac{x\pi }{7}) \\
\frac{1}{\theta } &=&2\frac{\left( s(1)s(2)-s(3)\right) }{1+\frac{1}{2}\frac{%
s^{2}(1)}{s(3)}\left( s(1)s(2)-s(3)\right) \left[ 1+\frac{s(3)}{s^{2}(1)}%
\right] } \\
\frac{A}{\theta } &=&4\frac{s^{2}(3)}{s(1)}\frac{1}{1+\frac{1}{2}\frac{%
s^{2}(1)}{s(3)}\left( s(1)s(2)-s(3)\right) \left[ 1+\frac{s(3)}{s^{2}(1)}%
\right] } \\
\frac{B}{\theta } &=&2\frac{s(3)}{s^{2}(1)}\frac{\left( s(1)s(2)-s(3)\right) 
}{1+\frac{1}{2}\frac{s^{2}(1)}{s(3)}\left( s(1)s(2)-s(3)\right) \left[ 1+%
\frac{s(3)}{s^{2}(1)}\right] } \\
\frac{\theta _{1}}{\theta } &=&\frac{64}{9}\frac{s^{2}(3)}{s^{2}(1)}\frac{1+%
\frac{1}{2}\frac{1}{s^{2}(2)}\left( s(1)s(2)-s(3)\right) \left[ \frac{%
s^{2}(3)}{s^{2}(1)s^{2}(2)}+\frac{s(3)}{s^{2}(1)}\right] }{1+\frac{1}{2}%
\frac{s^{2}(1)}{s(3)}\left( s(1)s(2)-s(3)\right) \left[ 1+\frac{s(3)}{%
s^{2}(1)}\right] } \\
\frac{\theta _{2}}{\theta } &=&\frac{s^{2}(2)}{s^{2}(1)}\frac{1+\frac{1}{2}%
\frac{s^{2}(1)}{s^{2}(3)}\left( s(1)s(2)-s(3)\right) \left[ \frac{s^{2}(3)}{%
s^{2}(1)s^{2}(2)}+\frac{s(3)}{s^{2}(1)}\right] }{1+\frac{1}{2}\frac{s^{2}(1)%
}{s(3)}\left( s(1)s(2)-s(3)\right) \left[ 1+\frac{s(3)}{s^{2}(1)}\right] }
\end{eqnarray*}

The calculated structure constants are universal quantities, the obtained
results can be exploited in the study of materials classified in the
universal class of the Tricritical 3-states Potts model. Also, one can
realize a Monte Carlo simulation program or study the finite size scaling to
compare with these results \cite{mccb-priv}, but one may note that the
operator content of the critical model contain several fields (15 parameters
). Consequently, there are many correlators to measure and it will be
probably difficult to do such a work, although there is the possibility to
verify certain results.

\begin{acknowledgement}
The authors would like to thank especially Pr. John McCabe for a very useful
discussions, \ for informing them about the existing simulation methods and
for the carefully examination of the manuscript.
\end{acknowledgement}

\end{document}